\documentclass[showpacs,10pt,twocolumn,prl]{revtex4-1}

\usepackage{amsmath}
\usepackage{amssymb}
\usepackage{graphics}
\usepackage{epsfig}
\usepackage{CJK}
\usepackage{color}
\usepackage{epstopdf}
\begin{document}

\begin{CJK*}{GBK}{Song}

\title{Surface Conductivity in Antiferromagnetic Semiconductor CrSb$_2$}
\author{Qianheng Du,$^{1,2,\dag}$ Huixia Fu,$^{3}$ Junzhang Ma, $^{4,5}$ A. Chikina,$^{4,*}$ M. Radovic,$^{4}$ Binghai Yan,$^{3}$ and C. Petrovic$^{1,2,\ddag}$}
\affiliation{$^{1}$Condensed Matter Physics and Materials Science Department, Brookhaven
National Laboratory, Upton, New York 11973, USA\\
$^{2}$Department of Materials Science and Chemical Engineering, Stony Brook University, Stony Brook, New York 11790, USA\\
$^{3}$3Department of Condensed Matter Physics, Weizmann Institute of Science, Rehovot 7610001, Israel\\
$^{4}$Swiss Light Source, Paul Scherrer Institute, CH-5232 Villigen PSI, Switzerland\\
$^{5}$Institute of Condensed Matter Physics, \'Ecole Polytechnique F\'ed\'erale de Lausanne, CH-10 15 Lausanne, Switzerland}

\date{\today}

\begin{abstract}
The contribution of bulk and surface to the electrical resistance along crystallographic \textit{b}- and \textit{c}-axes as a function of crystal thickness gives evidence for a temperature independent surface states in an antiferromagnetic narrow-gap semiconductor CrSb$_{2}$. ARPES shows a clear electron-like pocket at $\Gamma$-$Z$ direction which is absent in the bulk band structure. First-principles calculations also confirm  the existence of metallic surface states inside the bulk gap. Whereas combined experimental probes point to enhanced surface conduction similar to topological insulators, surface states are trivial since CrSb$_2$ exhibits no band inversion.
\end{abstract}

\maketitle

\end{CJK*}

Topological states on surfaces of topological insulators (TIs) are of high interest in quantum information and spintronics alike \cite{MooreJ,SmejkalL}. Such conducting states are immune to backscattering-induced localization and exhibit high mobiity and electron diffusion length \cite{FuL,PesinD}. Moreover, they also show efficient spin filtering, strong spin-momentum locking and highly efficient and Fermi level-dependent charge to spin current conversion \cite{WuJ,LiC,ShiomiY,KondouK}. Topological surface states in correlated electron materials were theoretically predicted within Topological Kondo Insulator (TKI) framework and experimentally verified on surfaces of SmB$_{6}$ crystals \cite{DzeroM,ZhangX,KimD,Wolgast,Syers}. TKI arise when bulk insulating gap opens due to hybridization of 4\textit{f} with conduction electron orbitals of different parity via band inversion mechanism at high symmetry point at the Brillouen zone and are embodiment of interacting topological phases of matter \cite{RachelS}. Kondo Insulator physics with reduced Coulomb repulsion has also been proposed for FeSi by Aeppli and Fisk \cite{FiskAeppli}. This was supported by neutron scattering and thermodynamic measurements \cite{Mason,Mandrus}. Interestingly, conducting surface states have also been observed in FeSi \cite{FangY}.

FeSb$_2$ and CrSb$_2$ crystallize in identical marcasite crystal structure and both are FeSi-like narrow gap semiconductors with dominant 3$d$ character of the electronic states near the valence- and conduction-band edges \cite{Cedomir2,Kuhn,Sales2,KoyamaT}. Whereas former compound features temperature-induced paramagnetism \cite{Cedomir2,ZaliznyakI}, the latter hosts relatively high-temperature antiferromagnetic (AFM) order below $T_{N}$ = 273 K and quasi-1D magnons \cite{Sales2,StoneM}. The relation to Kondo Insulator physics and surface conducting states have been discussed in FeSb$_2$ \cite{Cedomir2,Bentien2,PerucchiA,HongS,ChikinaA,XuKJ}. In this article we demonstrate the existence of the conducting states on CrSb$_{2}$; in contrast to SmB$_6$, topological states are trivial and are formed by the Cr 3$d$ orbitals.

Single crystals of CrSb$_2$ were grown as described previously \cite{DuQ}. Electrical transport was measured in a Quantum Design PPMS-9. Resistivity was measured by a standard four-probe method. Hall effect was measured with current along the \textit{b}-axis and magnetic field along the \textit{a}-axis. ARPES data were taken on in-situ cleaved crystals along the \textit{ac}-plane at PSI SIS beamline. The vaccum was better than $5\times10^{-11}$ mbar throughout the measurements.

\begin{figure}
\centerline{\includegraphics[scale=0.30]{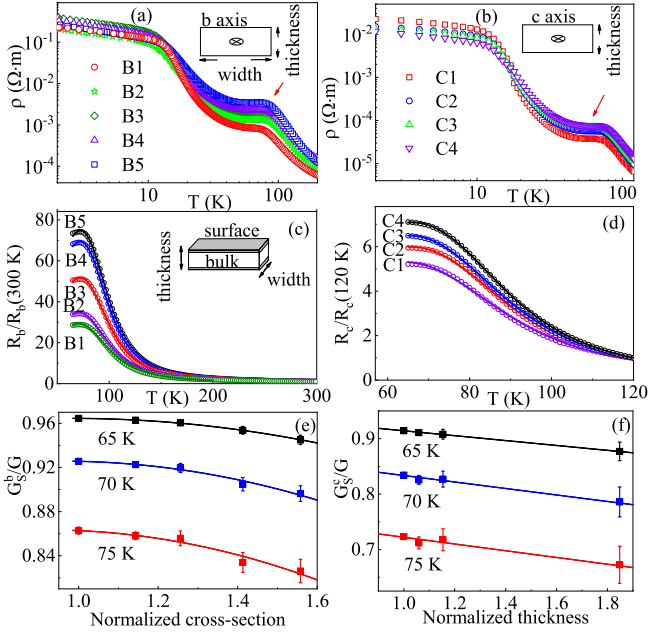}}
\caption{In both directions, $R(T)$ is well described by a thermally activated semiconducting bulk contribution with a surface contribution.Size dependence of the electrical resistivity for current flow along the \textit{b}-axis (a) and \textit{c}-axis (b). Resistivity along both axes have a size-dependent plateau from 50 K to 80 K (red arrows). The insets show the definition of sample dimensions. Temperature dependence of the normalized electrical resistance along \textit{b}-axis (c) and \textit{c}-axis (d). Solid lines represent fitting using two-channel conductance model which gives the conductivity for each conducting channel. (e,f) show the relationship between the normalized crystal sizes and the ratio of surface contribution to conductance along \textit{b}- and \textit{c}-axes at different temperatures.}
\end{figure}

Figure 1(a,b) presents temperature dependence of the resistivity $\rho(T)$ for a single crystal as a function of size reduction; crystal was oriented using a Laue camera and cut along \textit{b}- and \textit{c}-axis for resistivity measurement. In order to study the size-dependent resistivity, a bar-shaped sample was cut from a big single crystal along specific axes and sample size was varied by polishing. For the current path along the \textit{b}-axis crystal was reduced along both orthogonal directions in five steps B1-B5 as the cross-section decreases whereas for the \textit{c}-axis current path crystal cross-section was reduced in a single orthogonal direction in four steps C1-C4 [Fig. 1 inset]. There is a decrease in $\rho(T)$ values down to 20 K, as expected for a semiconductor in all investigated samples. For B1-B5 crystals $\rho(T)$ increases about 5 times for the sample size decrease from B1 (360 $\mu$m $\times$ 750 $\mu$m) to B5 (330 $\mu$m $\times$ 525 $\mu$m) whereas $\rho(T)$ doubles from C1 (400 $\mu$m) to C4 (216 $\mu$m). Below 20 K there is no monotonic increase; the $c$-axis resistivity shows a decrease in $\rho(T)$ with size reduction.

First we discuss high-temperature behavior. To eliminate uncertainty in the geometric factor arising from varying contact geometry, we plot the resistance ratio $R_b/R_b(300 K)$ and $R_c/R_c(120 K)$ for crystals with current path along \textit{b}- and \textit{c}-axis respectively, i.e. the $R(T)$ normalized to the resistance values at 300 K and 120 K respectively [Fig. 1(c,d)]. The resistance curves show similar qualitative behavior. The change from a high temperature thermal activated behavior to a plateau in $R(T)$ around (80 -100) K has been attributed to the strong electron-phonon interaction \cite{Sales2}. This is inconsistent with the thickness-dependence of electrical resistance presented in Fig. 1(c,d) which shows a clear separation of normalized resistance curves from a single trace at higher temperatures to distinct plateau values for each thickness at that temperature.

A simple parallel conductance model, with total conductance described by $G = G_S + G_B$, is used to extract the contribution from bulk and surface. Here, $G_S = 1/R_S$ is the surface contribution, which is assumed to be independent on $T$. The $G_B = 1/R_B$ is the bulk contribution-assumed to be thermally activated due to a bulk energy gap $\Delta$. Therefore, $R_S = 1/G_S$ and $R_B = 1/G_B \propto e^{\Delta/2k_BT}$ where $k_B$ is Boltzmann constant are geometry-dependent resistance. Then $R_b/R_b(300 K)$ and $R_c/R_c(120 K)$ are dimensionless and size-independent resistance ratio,

\begin{equation}
\begin{aligned}
r = [\frac{R(T)}{R(T_0)}]^{-1} = [r_S]^{-1} + [r_Be^{\frac{\Delta}{k_BT}}]^{-1}
\end{aligned}
\end{equation}

where $r_S \equiv R_S/R(T_0)$ and $r_B \equiv R_B/R(T_0)$ are the dimensionless, normalized surface and bulk resistance ratios ($T_0$ = 300 K and 120 K for b- and c-axis, respectively) \cite{Syers}.

Fitting results to this model using $r_s$, $r_B$ and $\Delta$ as free parameters are shown as solid lines in Fig. 1(c,d). We obtain a size-independent energy gap of 101.9(1) $\pm$ 0.4 meV, consistent with our electrical transport measurement and former report \cite{Sales2}. The calculated ratio of contribution from surface state $G_S/G$ using the fitted parameters for size reduction and current paths along \textit{b}- and \textit{c}-axes is presented in Fig. 1(e,f), respectively. The values show a clear relation to crystal sizes: For current path along the \textit{b}-axis where crystals size was reduced in two orthogonal directions, the ratio $G_S^b/G^b$ exhibits a quadratic dependence on the cross-section. For current path along the \textit{c}-axis where crystal size was varied along one orthogonal direction, the $G_S^c/G^c$ shows a linear trend with thickness change. This indicates decreasing relative contribution of the surface conductance relative to bulk conductance with increasing sample thickness. Conversely, at 65 K, the extrapolated values of $G_S/G \approx 1$ at the zero-thickness (or zero-cross-section) limit denote zero electrical conductance through bulk, as expected in the bulk-surface model. In the plateau region, the contribution of surface increases as temperature decreases. This confirms the presence of the surface state in the formation of $\rho(T)$ plateau.

ARPES measurements [Fig. 2(a)] do not indicate an obvious characteristic of electron-phonon coupling. Hence, it is unlikely that the electron-phonon interaction contributes to the formation of the plateau in temperatuare-dependent resistivity. There is also a clear electron-like pocket along the $\Gamma$$_0$-Z$_0$ direction. This state is absent in the bulk band structure calculations \cite{Kuhn}. It shows two-dimensional character on the corresponding Fermi surface map and appears in the gap of the bulk band structure, which is consistent with the surface nature of this state. The fitted value of electron effective mass is $2.18m_e$, where $m_e$ is the electron mass. The corresponding carrier density and Fermi velocity are 1.59$\times$10$^{17}$ m$^{-2}$ and 2.84$\times$10$^5$ m/s, respectively.

Hall effect offers further insight into the surface contribution to electronic transport. We used three crystals with different thickness: S1 (450 $\mu$m), S2 (345 $\mu$m) and S3 (156 $\mu$m). Hall resistivity $\rho_{xy}$ at 20 K [Fig. 2(b)] shows a transition from linear one-band behavior (S1) to two-bands behavior (S2 and S3) with thickness reduction. Figure 2(c) shows the Hall coefficient $R_H$ ($=\rho_{xy}/B$) for S1 and the high-field $R_H$ for S2 and S3. As the high-field limit of $R_H$ is determined only by the number and type of carriers, these can be used to estimate the apparent carrier concentration $10^{25} m^{-3}$. Below 20 K, Hall coefficient $R_H$ are similar for all crystals, whereas above 20 K Hall coefficients follow similar trend but the values for S1, S2 and S3 are different.

All experimental observations above indicate the increase in contribution from the surface states as the thickness decreases. The surface electron concentration in ARPES measurement is 1.59$\times$10$^{17}$ m$^{-2}$. This corresponds to effective 3D Hall coefficients of $-1.76\times10^{-4} m^3/C$ (S1), $-1.35\times10^{-4} m^3/C$ (S2) and $-6.13\times10^{-5} m^3/C$ (S3), much larger than the observed value. Hence, both bulk and surface electronic states account for the $R_H$ in S2 and S3 whereas thick S1 crystal shows single-band behavior since the signal from surface electrons is dwarfed by bulk electronic states. The two band electronic transport model

\begin{equation}
\begin{aligned}
\rho_{xy} = \frac{(R_s\rho_n^2+R_n\rho_s^2)B+R_sR_n(R_s+R_n)B^3}{(\rho_s+\rho_n)^2+(R_s+R_n)^2B^2}
\end{aligned}
\end{equation}

fits the whole $\rho_{xy}(B)$ curve well [Fig. 2(b)] \cite{Zhi}. Here, $R_n$ and $\rho_n$ are the Hall coefficient and resistivity of the bulk state. $R_s=t/(eN_s)$ and $\rho_s=\rho_{\Box}t$ are the Hall coefficient and resistivity of the surface state with $\rho_{\Box}$ the surface sheet resistance and $t$ the sample thickness. For S2, the fitting results are the surface mobility $\mu_s$ = 0.12 m$^2$/Vs and the bulk mobility $\mu_n$ = 7$\times$10$^{-4}$ m$^2$/Vs, along with the surface carrier concentration $N_s$ = 1.59$\times$10$^{17}$ m$^{-2}$ and the bulk carrier concentration $n$ = 3.7$\times$10$^{25}$ m$^{-3}$. For S3, the corresponding values are $\mu_s$ = 0.11 cm$^2$/Vs, $\mu_n$ = 1$\times$10$^{-3}$ m$^2$/Vs, $N_s$ = 1.59$\times$10$^{17}$ m$^{-2}$ and $n$=4.8$\times$10$^{25}$ m$^{-3}$, which are quite similar to the results of S2. This confirms the existence of thickness-independent surface state.

\begin{figure}
\includegraphics[scale=0.3]{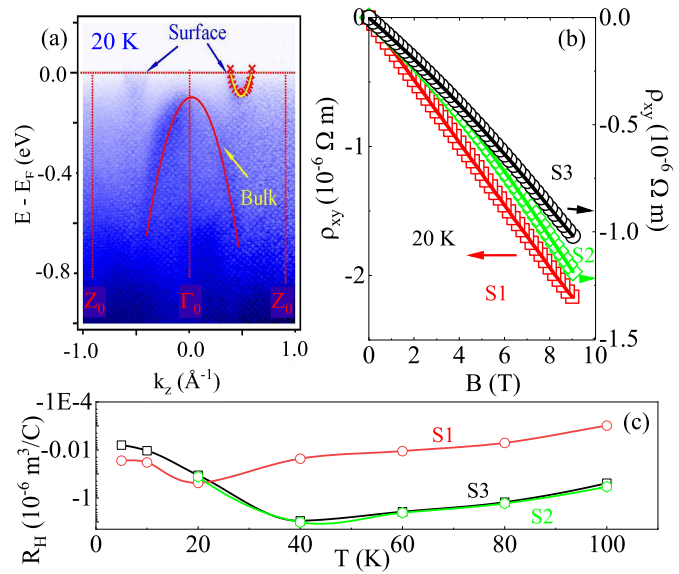}
\caption{(a) ARPES spectra at $\Gamma$$_0$-$Z$$_0$ cut in Brillouin zone measured at 20 K show an electron pocket which is absent in the bulk band structure calculations. (b) Hall resistivity at 20 K for different samples. Solid lines are fittings. (c) Hall effect measurements show a transition from one-band to two-bands behavior with decrease sample sizes, which confirms the increase contribution from the surface state with decrease sample size. Temperature dependence of the $R_H$ (S1) and high-field $R_H$ (S2 and S3).}
\end{figure}

In order to analyze the individual conductivity of the surface and bulk states we consider conductivity tensor $\sigma = \rho^{-1}$ , where $\sigma$ is the sum of surface and bulk contributions \cite{Ando}:

\begin{equation}
\begin{aligned}
\sigma_{xx} = \frac{\sigma_{xx}^s}{t}+\sigma_{xx}^b
\end{aligned}
\end{equation}

and $t$ is the crystal thickness \cite{Zhi}. From the Drude model,

\begin{equation}
\begin{aligned}
\sigma_{xx} = \frac{2n_s|e|}{t}\frac{\mu_s(T)}{1+\mu_s^2(T)B^2}+n_b(T)|e|\frac{\mu_b}{1+\mu_b^2B^2}
\end{aligned}
\end{equation}

where $e$ is the electron charge and $n$ and $\mu$ are carrier density and mobility, respectively. We use the subscript $s$ and $b$ to denote the surface and bulk contributions. According to the Matthiessen's rule \cite{Ashcroft} for the electron mobility, the $\mu_s$ is treated as $\frac{1}{\mu_s(T)}=\frac{1}{\mu_{s0}}(1+cT^{\gamma})$. The possible temperature dependence of $\mu_b$ was neglected since the thermal activation of $n_b(T)=n_{b0}exp(-\Delta_{bt}/T)$ dominates. The $n_s$, $\mu_{s0}$, $c$, $\gamma$, $n_{b0}$, $\Delta_{bt}$ and $\mu_b$ are free parameters in a fit to this two-band model of electrical conductivity.

Two crystals were polished to four different samples labeled as $R_{ij}$ where $i=1,2$ represents crystal number and $j=1,2$ represents different thickness. Larger j means thinner sample. The low temperature conductivity of these samples are shown in Fig. 3(a). The solid lines represent the fit of two-band conductivity model.The two-band model explains the low temperature conductivity well. From the fitting results, we calculated the surface and bulk contributions [Fig. 3(b)]. The conductivity of the bulk is nearly identical for all samples which is associated with similar amount of defects and imperfections. The contribution of surface state is independent of the temperature, which also confirms the validity of the model used to analyze resistivity at higher temperatures. It is of interest to note that there is a crossover in the conductivity around 15 K. This crossover explains the low temperature behavior shown in Fig. 1(b). Below the crossover temperature, the conductivity of surface state is higher than that of bulk state. As the sample thickness decreases, the contribution of surface state increases and the total conductivity increases. Above the crossover temperature the trend is opposite due to higher conductivity of bulk states.

CrSb$_{2}$ surface states feature enhanced effective mass over the bare electron mass and also smaller surface mobility when compared to canonical topological insulators \cite{Barriga,QuD1,QuD2}. However they are 5 - 20 times smaller when compared to massive surface states observed in SmB$_{6}$ \cite{Luo}. Electronic correlations and AFM order could play important role in the mass enhancement of the surface state; in that context it is of interest to note that conducting surface states have been observed in FeSi in transport measurements but ARPES data failed to detect such states possibly due to their location well above the Fermi level \cite{FangY,ChangdarS}.

\begin{figure}
\centerline{\includegraphics[scale=0.3]{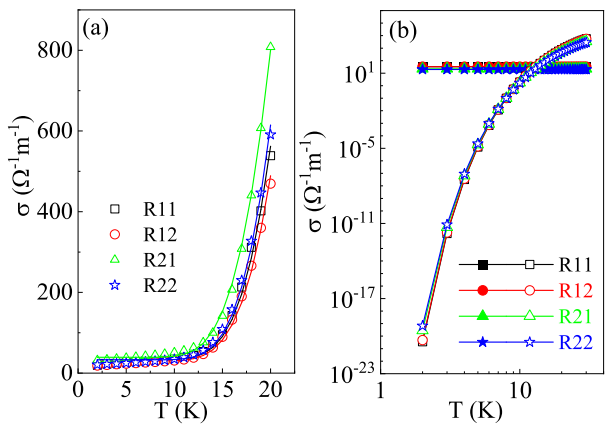}}
\caption{ (a) Conductivity vs temperature below 20 K. The sample labels are described in the text. Solid lines are two-band model fitting. (b) Conductivity of surface (solid) and bulk state (open symbols).
}
\end{figure}


We performed density-functional theory (DFT) calculations within the local density approximation \cite{Ceperley} using the VASP package \cite{Kresse}. We considered the experimentally reported AFM configuration \cite{Sales2} and obtained an insulating bulk band structure, similar to previous reports \cite{DuQ}. In the AFM bulk structure, the lattice parameters are $a$ = 6.008 ${\AA}$, $b$ = 13.726 ${\AA}$ and $c$ = 6.536 ${\AA}$. The bulk phase is likely a trivial insulator since we do not find clear topological feature.
To simulate the surface states, we constructed slab models for the ac, ab and bc planes. The slab models inherit the bulk AFM configuration. Then we calculated the surface band structures by including the spin-orbit coupling (SOC).

\begin{figure}
\centerline{\includegraphics[scale=0.30]{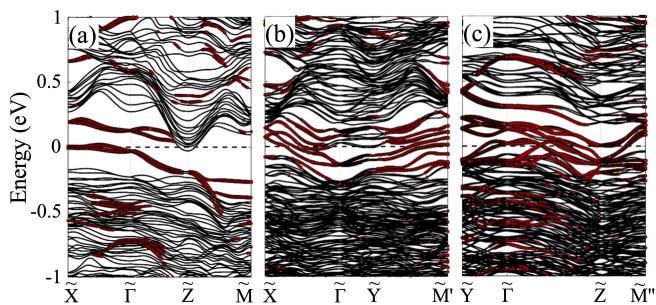}}
\caption{ Surface band structures. (a-c) Surface structures of CrSb$_2$ for $ac$, $ab$ and $bc$ planes, respectively. The red dots highlight the surface states. It corresponds to the AFM magnetic configuration. The Fermi energy is set to zero.
}
\end{figure}

Figure 4 shows the surface band structure. For different surfaces, there are metallic surface states inside the bulk gap, which exhibit strong SOC splitting. We take the $ac$ plane for example [Fig. 4(a)]. The $ac$ surface forms a chain-like structure along the $c$-axis. In the momentum space, surface bands follow this anisotropy and are more dispersive along the $\Gamma$ - $Z$ direction, compared to the $\Gamma$ - $X$ direction. Corresponding surface bands are dominantly contributed by the surface Cr - 3$d$ orbitals.
We note that slab models have atomically flat terminations. In reality, the surface atomic configuration may be strongly disordered, leading to the blurred surface bands, as shown in Fig. 2(a) along the $\Gamma$$_0$ - $Z$$_0$ line. When comparing the calculation and the ARPES data, it should be noted that the Brillouin zone is folded along $\Gamma$$_0$ - $Z$$_0$ direction in the calculation with the AFM phase. The surface states entered at $\tilde{Z}$ point is located between $\Gamma$$_0$ and Z$_0$ in Fig. 2(a); that indicates well agreement between the calculation and experiment. On the other hand, for the bulk band, the ARPES data do not show any band folding between $\Gamma$$_0$ and Z$_0$, most probably because the AFM induced magnetic filed is not strong enough to change the electronic structure dramatically. Thus, the spectral weight of the folded bands can be too weak to be observed. In brief, both calculations and ARPES show the existence of metallic surface states inside the bulk energy gap, which is consistent with the transport measurement.

In summary, we have presented the first evidence that CrSb$_{2}$ hosts surface conducting states. The thickness-dependent resistivity and the transition from one-band to two-band Hall effect come from the increased contribution from the surface state as the sample size decreases. The crossover in the conductivity of the bulk and surface states explains the plateau and low temperature behavior in resistivity. The surface states are also observed in ARPES measurement, in good agreement with electronic transport. First principle calculations indicate that surface states in the bulk energy gap are topologically trivial.

Work at Brookhaven is supported by the U.S. DOE under Contract No. DE-SC0012704. B.Y. acknowledges the financial support by the Willner Family Leadership Institute for the Weizmann Institute of Science, the
Benoziyo Endowment Fund for the Advancement of Science, Ruth and Herman Albert Scholars Program for New Scientists, and the European Research Council (ERC) under the European Union's Horizon 2020 research and innovation program (grant no. 815869). J.-Z.M. A. C and M. R. were supported by Project No. 200021-182695 funded by the Swiss National Science Foundation. ARPES experiments were conducted at the Surface/Interface Spectroscopy (SIS) beamline of thevSwiss Light Source at the Paul Scherrer Institut in Villigen, Switzerland. The authors thank the technical staff at the SIS beamline for their support.

$^{*}$Present address: Department of Physics and Astronomy, Aarhus University, Aarhus 8000 Denmark
$^{\ddag}$petrovic@bnl.gov
$^{\dag}$qdu@bnl.gov

\end{document}